%% file: acl_latex.tex
\pdfoutput=1

\documentclass[11pt]{article}

\usepackage[final]{acl}

\usepackage{times}
\usepackage{latexsym}

\usepackage[T1]{fontenc}

\usepackage[utf8]{inputenc}

\usepackage{microtype}

\usepackage{inconsolata}

\usepackage{graphicx}
\usepackage{amsmath}
\usepackage{tcolorbox}
\usepackage{booktabs}
\usepackage{times}         
\usepackage{latexsym}      
\usepackage{microtype}     
\usepackage{graphicx}      
\usepackage{subfigure}     
\usepackage{booktabs}      
\usepackage{hyperref}      
\usepackage{amsmath}       
\usepackage{amssymb}       
\usepackage{mathtools}     
\usepackage{amsthm}        
\usepackage{algpseudocode} 
\usepackage{multirow}      
\usepackage{bbm}           
\usepackage{colortbl}      
\usepackage{xspace}        
\usepackage{wrapfig}       
\usepackage{arydshln}      
\usepackage{makecell}      
\usepackage{tcolorbox}     
\usepackage{listings}      
\usepackage{enumitem}      
\usepackage{url}           
\usepackage{amsfonts}      
\usepackage{nicefrac}      
\usepackage{xcolor}        
\usepackage{pifont}        
\usepackage{CJKutf8}       
\usepackage[utf8]{inputenc}

\title{CodeIF: Benchmarking the Instruction-Following Capabilities of Large Language Models for Code Generation}

\newcommand{\bench}{\textsc{CodeIF}}

\author{
 \textbf{Kaiwen Yan\textsuperscript{1 *}},
 \textbf{Hongcheng Guo\textsuperscript{1 * †}},
 \textbf{Xuanqing Shi\textsuperscript{2}}
 \\
 \textbf{Shaosheng Cao\textsuperscript{3}},
 \textbf{Donglin Di\textsuperscript{2}},
 \textbf{Zhoujun Li\textsuperscript{1}}
\\
\\
 \textsuperscript{1}Beihang University,
 \textsuperscript{2}Tsinghua University,
 \textsuperscript{3}Xiaohongshu
\\
 \small{
   {lin\_rany@foxmail.com},
   {hongchengguo@buaa.edu.cn},
   {sxq23@mails.tsinghua.edu.cn}
   }
\\
\small
   {caoshaosheng@xiaohongshu.com},
   {donglin.ddl@gmail.com},
   {lizj@buaa.edu.cn}
 }

\begin{document}
\maketitle
\let\oldthefootnote\thefootnote 
\renewcommand{\thefootnote}{} 
\footnotetext{\textsuperscript{*}Equal contribution}
\footnotetext{\textsuperscript{\dag}Corresponding Author}
\let\thefootnote\oldthefootnote 

\begin{abstract}
With the rapid advancement of Large Language Models (LLMs), the demand for robust instruction-following capabilities in code generation tasks has grown significantly. Code generation not only facilitates faster prototyping and automated testing, but also augments developer efficiency through improved maintainability and reusability of code. In this paper, we introduce CodeIF, the first benchmark specifically designed to assess the abilities of LLMs to adhere to task-oriented instructions within diverse code generation scenarios. CodeIF encompasses a broad range of tasks, including function synthesis, error debugging, algorithmic refactoring, and code explanation, thereby providing a comprehensive suite to evaluate model performance across varying complexity levels and programming domains. We conduct extensive experiments with LLMs, analyzing their strengths and limitations in meeting the demands of these tasks. The experimental results offer valuable insights into how well current models align with human instructions, as well as the extent to which they can generate consistent, maintainable, and contextually relevant code. Our findings not only underscore the critical role that instruction-following LLMs can play in modern software development, but also illuminate pathways for future research aimed at enhancing their adaptability, reliability, and overall effectiveness in automated code generation. \footnote{CodeIF data and code are publicly available:
\url{https://github.com/lin-rany/codeIF}}.
\end{abstract}
\input{intro}

\input{CodeIF_overview}
\input{CodeIF_DataSet_building}

\input{metrics}

\input{experiment}
\input{analysis}

\input{related_work}
\input{conclusion}

\clearpage
\input{limitations}

\bibliography{custom}


\appendix
\onecolumn
\input{appendix}

\end{document}

%% file: intro.tex
\section{Introduction}


With the rapid advancement of large language models (LLMs), automated code generation is undergoing a profound transformation. While LLMs have demonstrated promising capabilities in programming tasks, their ability to comprehend and execute complex instructions remains a challenge~\cite{liu2024fullstackbenchevaluatingllms, zhang2023repocoder}. To drive progress in this field, a comprehensive and systematic evaluation framework is essential~\cite{jiang2024followbenchmultilevelfinegrainedconstraints, zhou2023instructionfollowingevaluationlargelanguage}.

This study introduces CodeIF, a benchmark designed to assess LLMs’ instruction-following capabilities in code generation. Built upon insights from existing evaluation sets like McEval~\cite{mceval} and FullStackBench~\cite{liu2024fullstackbenchevaluatingllms}, CodeIF is tailored for multi-language environments, covering Java, Python, Go, and C++. It categorizes tasks by difficulty and systematically evaluates models across 50 fine-grained sub-instructions, providing a nuanced understanding of their strengths and weaknesses.

To ensure rigorous assessment, we propose four novel evaluation metrics: Completely Satisfaction Rate (CSR), Soft Satisfaction Rate (SSR), Rigorous Satisfaction Rate (RSR), and Consistent Continuity Satisfaction Rate (CCSR). These metrics measure models' ability to handle multi-constraint problems from different perspectives, including full compliance, average constraint satisfaction, logical coherence, and consistency in instruction execution. By offering a structured evaluation framework, CodeIF provides valuable insights into the current state and future direction of LLM-driven code generation.
Overall, our contributions are mainly four-fold:
\begin{enumerate}[leftmargin=*,topsep=2pt,itemsep=2pt,parsep=0pt]
    \item \textbf{Innovative Benchmark.} We introduce \textbf{CodeIF}, the first systematic benchmark for evaluating LLMs' instruction-following capabilities in code generation. CodeIF categorizes tasks into \textbf{8 main types and 50 fine-grained sub-instructions}, ensuring a comprehensive assessment of model performance.
    
    \item \textbf{Automated High-Quality Instruction Generation.} Leveraging advanced LLMs like GPT-4, we develop a method to automatically generate constraint-based instruction lists. This approach enhances evaluation depth by incorporating instructional dependencies while minimizing human intervention.
    
    \item \textbf{Novel Evaluation Metrics.} We propose a new framework with four key metrics (\textbf{CSR, SSR, RSR, and CCSR}) tailored for code generation tasks. These metrics assess models' ability to handle multi-constraint problems across different dimensions, offering deeper insights and new benchmarks for future research.
    
    \item \textbf{Extensive Evaluation and Analysis.} We systematically evaluate \textbf{35 state-of-the-art LLMs}, including both open-source and commercial models, across multiple programming languages and difficulty levels. Our experiments uncover current strengths and limitations, providing clear directions for future advancements.
\end{enumerate}

%% file: CodeIF_overview.tex
\section{\bench{}}

\paragraph{Overview:}
\begin{figure}
    \centering
    \includegraphics[width=0.5\textwidth]{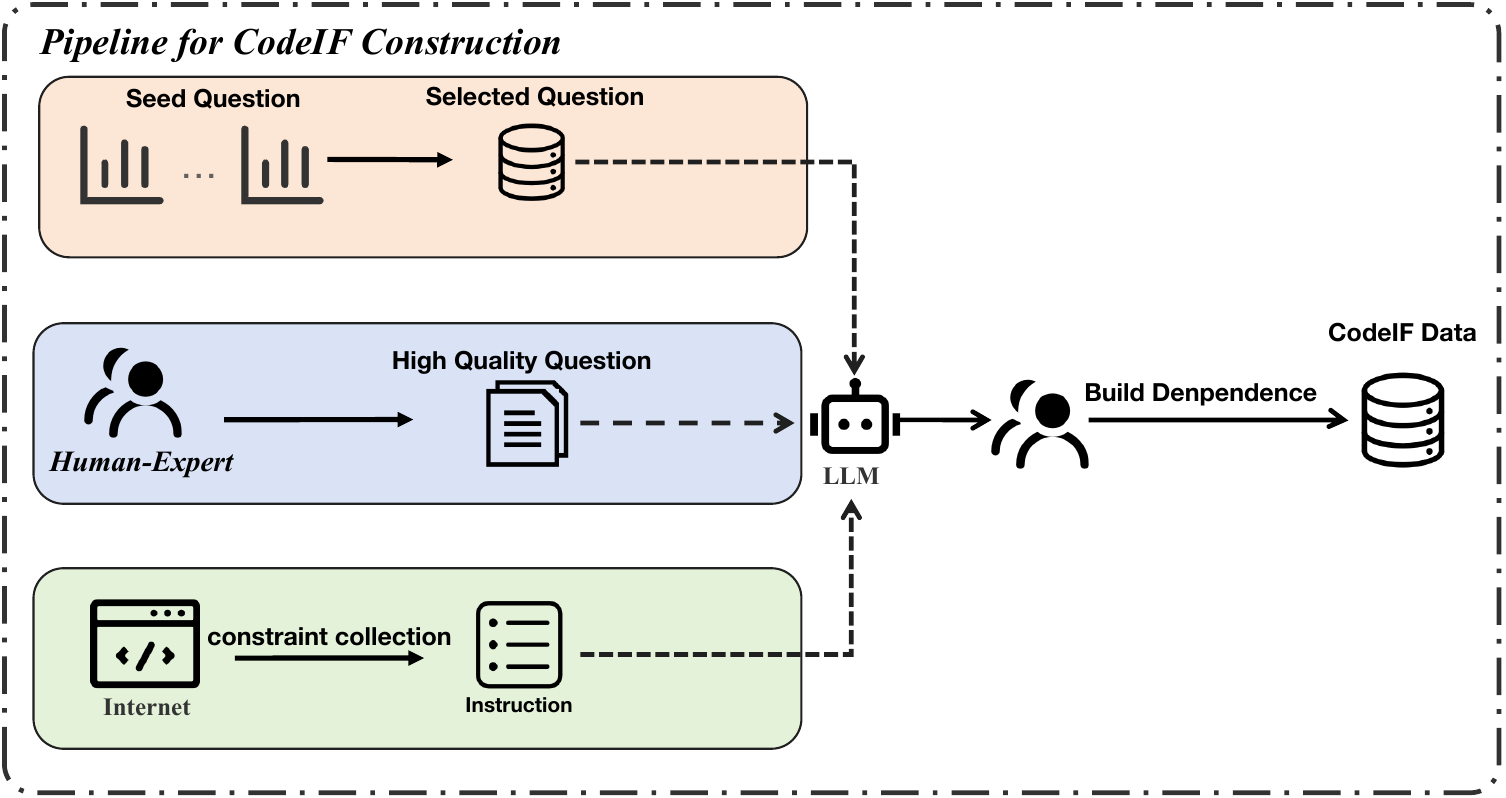}
    \caption{ The construction process of CodeIF. The first step involves the construction of constraint instructions, followed by the assembly of the dataset, and finally the construction of dependencies between instructions.}
    \label{fig:code_IF_all}
\end{figure}

 As shown in Figure~\ref{fig:code_IF_all}, CodeIF is constructed by collecting and refining constraint instructions from real code generation tasks, then combining these tasks with LLM outputs and human review to create a high-quality evaluation dataset.
 

%% file: CodeIF_DataSet_building.tex
\subsection{Building}
The construction of the \bench{} dataset involves two phases: collecting constraint instructions (Section~\ref{sec:Constraint_Instructions_Collection}) and processing data to create the final CodeIF evaluation dataset (Section~\ref{sec:Data_Construction}).

\subsection{Constraint Instructions Collection}
\label{sec:Constraint_Instructions_Collection}

The first phase of our work centers on code generation, constructing the \textbf{CodeIF} evaluation dataset through two steps: (1) collecting and verifying constraint instructions, and (2) using them for dataset generation.

We analyze benchmarks like \textbf{McEval}~\cite{mceval} and \textbf{FullStackBench}~\cite{liu2024fullstackbenchevaluatingllms} to develop an instruction system spanning \textbf{eight categories}, each targeting specific aspects of code generation for a fine-grained assessment of LLMs' instruction-following abilities. Constraints are decomposed into \textbf{atomic instructions} with explicit directives, enabling objective binary evaluation (yes/no) and minimizing subjective interpretation. The eight categories cover both \textbf{architectural-level specifications} and \textbf{variable-level implementation controls}, ensuring comprehensive constraint coverage. Specifically, the \textbf{Global} category evaluates adherence to overarching specifications, while \textbf{Structural Control} focuses on control structures (e.g., loops, conditionals) and data structures. \textbf{Variable} constraints assess naming and initialization. Higher abstraction levels include \textbf{Interface, Function,} and \textbf{Class} constraints for program components, while the \textbf{File} category tests cross-file dependencies and external library handling. The \textbf{Combination} category integrates constraints across dimensions, challenging models with complex scenarios.

\begin{figure}
    \centering
    \includegraphics[width=0.8\linewidth]{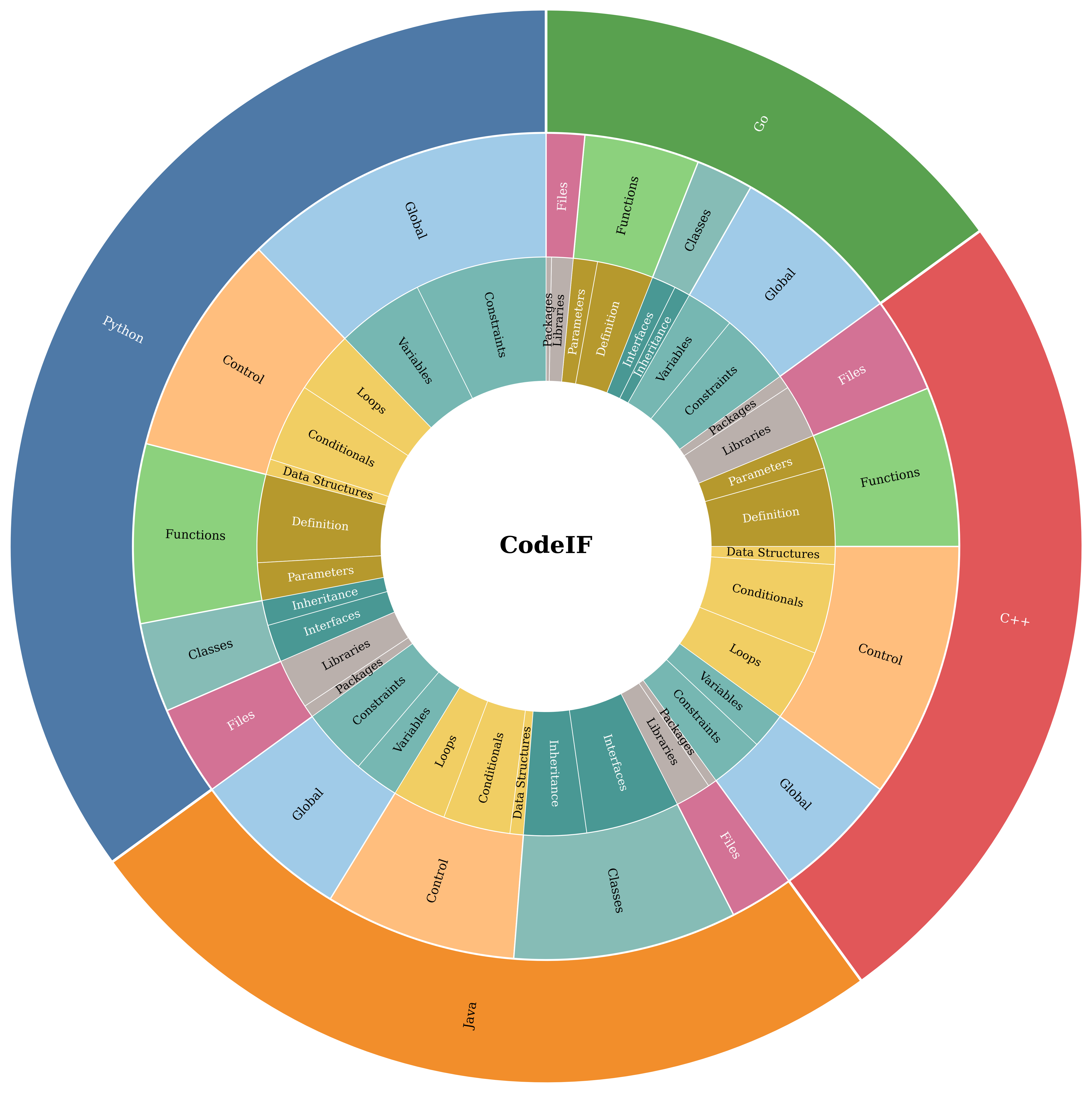}
    \caption{ CodeIF Constraints Instruction Distribution}
    \label{fig:instruction_analysis}
\end{figure}

Figure~\ref{fig:instruction_analysis} presents CodeIF's distribution across programming languages and categories. The evaluation system features \textbf{8 categories} and \textbf{50 fine-grained constraint instructions}, systematically assessing LLMs' code generation performance. By organizing constraints clearly, the system identifies strengths and weaknesses, guides optimization, and advances automated code generation. The full list of constraints is in Appendix~\ref{tab:Constraint_Instruction_Tabl}.

\subsection{Data Construction}
\label{sec:Data_Construction}

\paragraph{Multi-Language and Difficulty-Differentiated Benchmark Design}
To ensure diversity and comprehensiveness in evaluation, we carefully selected code generation tasks across four mainstream programming languages—Java, Python, Go, and C++—from leading benchmarks such as \textbf{McEval}~\cite{mceval} and \textbf{FullStackBench}~\cite{liu2024fullstackbenchevaluatingllms}. These languages, spanning both dynamic and static paradigms, create a rich linguistic environment that enhances multi-language assessment.

To further refine the evaluation, we categorize tasks into two difficulty levels: \textbf{Easy} and \textbf{Hard}. The \textbf{Hard} set includes longer, more intricate instruction lists, designed to rigorously test LLMs’ ability to handle complex constraints.

\paragraph{Automated Generation of Constraint Instructions}
We used large language models (LLMs) like \textbf{GPT-4} to create task-specific instruction lists for code generation tasks. We prepared 20 detailed examples and formulated concise atomic instructions for accuracy. These examples guided LLMs in refining tasks and streamlining instructions to enhance clarity and output quality.

\begin{figure}
    \centering
    \includegraphics[width=.4\textwidth]{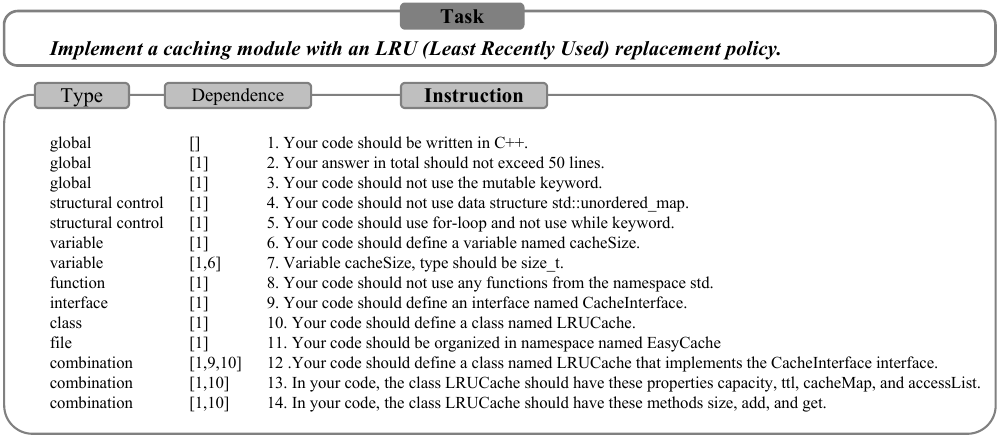}
    \caption{  Specific cases of the CodeIF dataset, 'Task' denotes the specific generation task, 'Type' refers to the type of constraint, and 'Dependence' indicates the prerequisite constraints for this constraint.}
    \label{fig:CodeIF_case}
\end{figure}

\paragraph{Constructing Instruction Dependencies}
We utilized LLMs to map dependencies between atomic constraints, improving our evaluation framework’s precision and verification accuracy. By understanding the dependencies among instructions, we outlined clear steps for tasks like function creation, which involve naming the function, defining parameter types, and coding the body. Incorporating these dependencies enhances our evaluation system, more accurately assessing the model’s capability with complex instructions and identifying areas for improvement. Figure~\ref{fig:CodeIF_case} illustrates a CodeIF task with its instruction sequence and dependencies.

\subsection{Data Analysis}


\begin{table}[!t]
\centering
\scriptsize
\renewcommand{\arraystretch}{1} 
\resizebox{\linewidth}{!}{ 
\begin{tabular}{c|c|c|c|c|c|c}
\hline
Set & Num & Avg.Instr & Go & Python & Java & C++ \\
\hline
Easy  & 600  & 11.99 & 127 & 165 & 176 & 132 \\
\hline
Hard  & 600  & 13.80 & 103 & 183 & 177 & 137 \\
\hline
Full  & 1200 & 12.90 & 230 & 348 & 353 & 269 \\
\hline
\end{tabular}
}
\caption{CodeIF dataset statistics, showing the statistical information of different difficulty classifications. Avg.Instr represents the average length of the atomic constraint instruction list.}
\label{tab:dataset_stats}
\end{table}

\paragraph{ CodeIF Static Analysis}
\begin{figure}
    \centering
    \includegraphics[width=.4\textwidth]{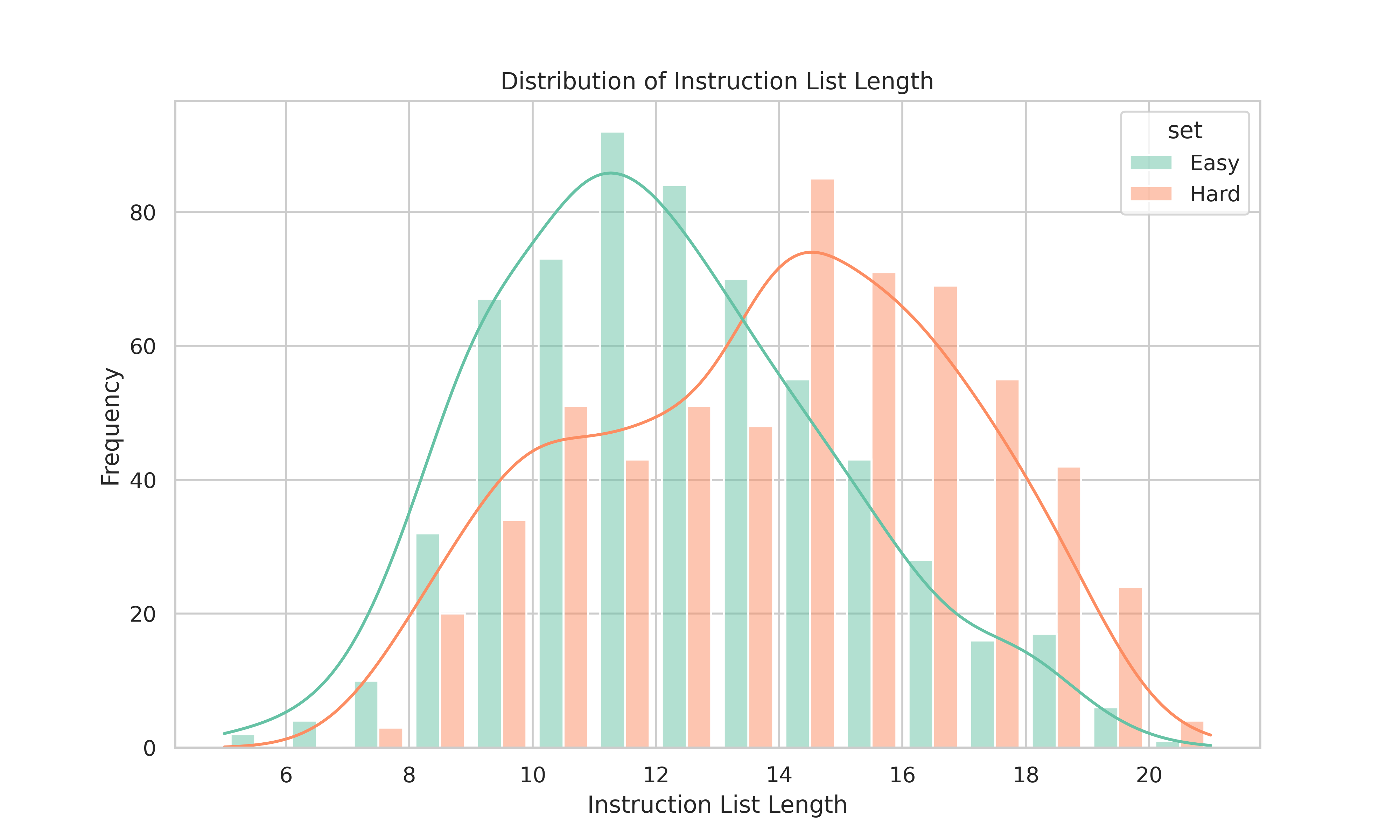}
    \caption{ The distribution of atomic instruction list lengths in datasets of different difficulties.}
    \label{fig:instruction_len_distribution_v2}
\end{figure}


Table~\ref{tab:dataset_stats} categorizes the dataset into three difficulty levels: \textbf{Easy}, \textbf{Hard}, and \textbf{Full}. Both Easy and Hard sets contain 600 tasks, while the Full dataset combines them, totaling 1,200 tasks across Go, Python, Java, and C++. \textbf{Java} has the most tasks (353), followed by \textbf{Python} (348), \textbf{C++} (269), and \textbf{Go} (230). The Easy set averages \textbf{11.99} instructions per task, the Hard set \textbf{13.8}, and the Full dataset \textbf{12.9}, reflecting increasing complexity. Figure~\ref{fig:instruction_len_distribution_v2} shows task length distribution.

\begin{table*}[!t]
\centering
\scriptsize
\renewcommand{\arraystretch}{1} 
\resizebox{\textwidth}{!}{ 
\begin{tabular}{c|c|c|c|c|c|c|c|c}
\hline
Set & Global & Structural Control & Variable & Interface & Function & Class & File & Combination \\
\hline
\textbf{Easy} & 1638 & 1008 & 1336 & 427 & 569 & 544 & 723 & 953 \\
\hline
\textbf{Hard} & 1890 & 1193 & 1479 & 505 & 659 & 623 & 802 & 1142 \\
\hline
\textbf{Full} & 3528 & 2201 & 2815 & 932 & 1228 & 1167 & 1525 & 2095 \\
\hline
\end{tabular}
}
\caption{CodeIF dataset statistics information, showing the distribution of atomic restriction instruction categories under different difficulty classifications.}
\label{tab:codeif_dataset_statistics_information}
\end{table*}

\paragraph{Constraint Instruction Analysis}

Table~\ref{tab:codeif_dataset_statistics_information}compares instruction distribution across difficulty levels. The \textbf{Hard} set consistently has more instructions per category than the \textbf{Easy} set, with the \textbf{Global} category averaging \textbf{2.5} instructions in Easy and over \textbf{3} in Hard. This indicates greater challenges for models as task complexity rises.
More analysis is in Appendix~\ref{app:dataanalysis}.


%% file: metrics.tex
\section{Metrics}

Ensuring that large language models (LLMs) accurately follow instructions is crucial for code generation. To precisely evaluate this capability, we introduce four novel metrics designed to assess how LLMs handle code generation tasks with multiple constraints: \textbf{Completely Satisfaction Rate (CSR)}, \textbf{Soft Satisfaction Rate (SSR)}, \textbf{Rigorous Satisfaction Rate (RSR)}, and \textbf{Consistent Continuity Satisfaction Rate (CCSR)}. These metrics provide a comprehensive evaluation from different perspectives.

For a dataset with $m$ problems, each problem contains a set of $n_i$ constraints. We define CSR and SSR as follows:

\paragraph{Completely Satisfaction Rate (CSR):}
\begin{equation}
\small
\text{CSR} = \frac{1}{m} \sum_{i=1}^{m} \left( \prod_{j=1}^{n_i} r_{i,j} \right)
\end{equation}
where $r_{i,j} \in [0,1]$ indicates whether the $j$-th constraint in the $i$-th problem is satisfied.

\paragraph{Soft Satisfaction Rate (SSR):}
\begin{equation}
\small
\text{SSR} = \frac{1}{m} \sum_{i=1}^{m} \left( \frac{\sum_{j=1}^{n_i} r_{i,j}}{n_i} \right)
\end{equation}
SSR evaluates the average proportion of constraints satisfied per problem, providing a more flexible assessment.

\paragraph{Rigorous Satisfaction Rate (RSR)}
In code generation, some constraints depend on prior instructions, particularly in \textbf{Combination} constraints. To account for dependencies, we define RSR as:
\begin{equation}
\small
\text{RSR} = \frac{1}{m} \sum_{i=1}^{m} \left( \frac{\sum_{j=1}^{n_i} \left[ r_{i,j} \cdot \prod_{k \in D_{i,j}} r_{i,k} \right]}{n_i} \right)
\end{equation}
where $D_{i,j}$ represents the set of constraints that the $j$-th constraint in the $i$-th problem depends on.

\paragraph{Consistent Continuity Satisfaction Rate (CCSR)}
In many code generation tasks, maintaining continuous adherence to instructions is essential. To measure this ability, we define CCSR as:
\begin{equation}
\scriptsize
\text{CCSR} = \frac{1}{m} \sum_{i=1}^{m} \frac{L_i}{n_i}, \\
L_i = \max \Bigl\{ l \,\Big|\, \exists t \mathbin{\in} [1, n_i{-}l{+}1],\ 
\prod_{\mathclap{j=t}}^{\mathclap{t+l-1}} r_{i,j} = 1 \Bigr\}
\end{equation}
where $L_i$ represents the longest consecutive sequence of satisfied constraints in problem $i$.

%% file: experiment.tex
\section{Experiment}

\subsection{Experimental Setup}
The temperature coefficient is set to 0 to ensure output determinism, with a maximum generation length of 4096 tokens. All other settings follow the official default parameters for each model. Commercial API models are accessed through the latest available interface as of December 2024. All experiments are conducted using the official API and 8 H800(80G).

\subsection{Automatic Evaluation}
To ensure robust evaluation, we used LLMs and human experts to verify model adherence to atomic constraints. Constraints were decomposed into atomic elements, enabling objective binary evaluations (\textit{Yes/No}) over subjective judgments. Following FairEval~\cite{faireval}, \textit{GPT-4-1106-Preview} was the primary evaluation tool (prompt details in Appendix~\ref{prompt:judge_prompt}). Three domain experts manually annotated 150 stratified samples. Statistical analysis showed strong agreement, with Pearson correlations of \textbf{0.87} (LLM-human) and \textbf{0.85} (inter-human), confirming high consistency. Baselines are in Appendix~\ref{app:baselines}.


\subsection{Main Experiments}

\begin{table*}[t]
\centering
\resizebox{1.0\textwidth}{!}{
\begin{tabular}{ccccccccccccc}
\toprule
\multirow{2}{*}{Models} & \multicolumn{3}{c}{CSR} & \multicolumn{3}{c}{SSR} & \multicolumn{3}{c}{RSR} & \multicolumn{3}{c}{CCSR} \\
\cmidrule(lr){2-13} 
 & Full & Easy & Hard & Full & Easy & Hard & Full & Easy & Hard & Full & Easy & Hard \\
\midrule
Llama-3.2-1B-Instruct & 0.034 & 0.046 & 0.022 & 0.218 & 0.277 & 0.159 & 0.182 & 0.231 & 0.132 & 0.152 & 0.197 & 0.107 \\
Llama-3.1-8B-Instruct & 0.145 & 0.187 & 0.102 & 0.467 & 0.544 & 0.388 & 0.418 & 0.493 & 0.340 & 0.370 & 0.444 & 0.295 \\
Qwen2.5-Coder-7B-Instruct & 0.142 & 0.198 & 0.087 & 0.514 & 0.590 & 0.438 & 0.453 & 0.533 & 0.373 & 0.390 & 0.463 & 0.318 \\
Qwen2.5-7B-Instruct & 0.153 & 0.201 & 0.104 & 0.535 & 0.599 & 0.471 & 0.475 & 0.546 & 0.405 & 0.416 & 0.479 & 0.353 \\
Ministral-8B & 0.161 & 0.205 & 0.116 & 0.552 & 0.614 & 0.489 & 0.486 & 0.552 & 0.419 & 0.431 & 0.490 & 0.371 \\
Gemma-2-9B-It & 0.171 & 0.210 & 0.131 & 0.573 & 0.642 & 0.504 & 0.513 & 0.587 & 0.440 & 0.445 & 0.508 & 0.383 \\
Qwen2.5-Coder-32B-Instruct & 0.365 & 0.422 & 0.307 & 0.736 & 0.767 & 0.704 & 0.679 & 0.723 & 0.635 & 0.634 & 0.669 & 0.599 \\
Gemma-2-27B-It & 0.245 & 0.300 & 0.190 & 0.658 & 0.709 & 0.607 & 0.596 & 0.652 & 0.540 & 0.533 & 0.588 & 0.478 \\
Qwen2.5-32B-Instruct & 0.294 & 0.337 & 0.251 & 0.680 & 0.722 & 0.638 & 0.621 & 0.674 & 0.568 & 0.560 & 0.604 & 0.515 \\
Qwen2.5-72B-Instruct & 0.281 & 0.319 & 0.244 & 0.685 & 0.734 & 0.634 & 0.621 & 0.677 & 0.564 & 0.569 & 0.619 & 0.518 \\
Llama-3.3-70B-Instruct & 0.307 & 0.359 & 0.255 & 0.698 & 0.749 & 0.647 & 0.632 & 0.691 & 0.574 & 0.589 & 0.643 & 0.536 \\
Gemini-Exp-1206 & 0.357 & 0.410 & 0.303 & 0.744 & 0.781 & 0.707 & 0.685 & 0.734 & 0.636 & 0.636 & 0.675 & 0.597 \\
GPT-4O-2024-11-20 & 0.383 & 0.441 & 0.325 & 0.748 & 0.792 & 0.702 & 0.689 & 0.745 & 0.633 & 0.650 & 0.698 & 0.602 \\
Claude-3-5-Sonnet-20241022 & \textbf{0.444} & \textbf{0.525} & \textbf{0.362} & 0.727 & 0.784 & 0.669 & 0.692 & 0.757 & 0.626 & 0.652 & 0.715 & 0.587 \\
Deepseek-V3 & 0.414 & 0.468 & 0.359 & \textbf{0.821} & \textbf{0.847} & \textbf{0.794} & \textbf{0.764} & \textbf{0.806} & \textbf{0.723} & \textbf{0.712} & \textbf{0.743} & \textbf{0.680} \\
\bottomrule
\end{tabular}}
\caption{CodeIF evaluation results of different difficulties. We use bold font to mark the best results in all models.}
\label{tab:code_if_evaluation_selected}
\end{table*}

Table~\ref{tab:code_if_evaluation_selected} evaluates CodeIF using four metrics: \textbf{CSR}, \textbf{SSR}, \textbf{RSR}, and \textbf{CCSR}. Detailed results are in Appendix~\ref{app:moreresults}.

\paragraph{Overview.}  
DeepSeek-V3 and Claude-3-5-Sonnet-20241022 lead across metrics, excelling in complex tasks. However, the highest \textbf{CSR} on Hard tasks is just \textbf{0.362}, showing challenges in meeting strict constraints.

\paragraph{Model Scale Trends.}  
Larger models generally perform better, as seen in Qwen2.5 series. However, the \textbf{Llama3} series shows inconsistent results, highlighting the importance of architecture, data quality, and optimization.

\paragraph{Open vs. Closed Models.}  
Closed-source models like GPT-4O and Claude-3-5 outperform open-source models, especially under complex constraints. While large open-source models (e.g., Qwen2.5-72B-Instruct) are competitive, they lag due to differences in data quality and RLHF techniques.

\paragraph{Task Difficulty Impact.}  
Performance drops with increasing task complexity. For instance, GPT-4O's \textbf{CSR} falls from \textbf{0.441} on Easy tasks to \textbf{0.325} on Hard tasks, highlighting the challenge of strict constraints.

%% file: analysis.tex
\section{In-Depth Analysis}

\begin{figure}
    \centering
    \includegraphics[width=.5\textwidth]{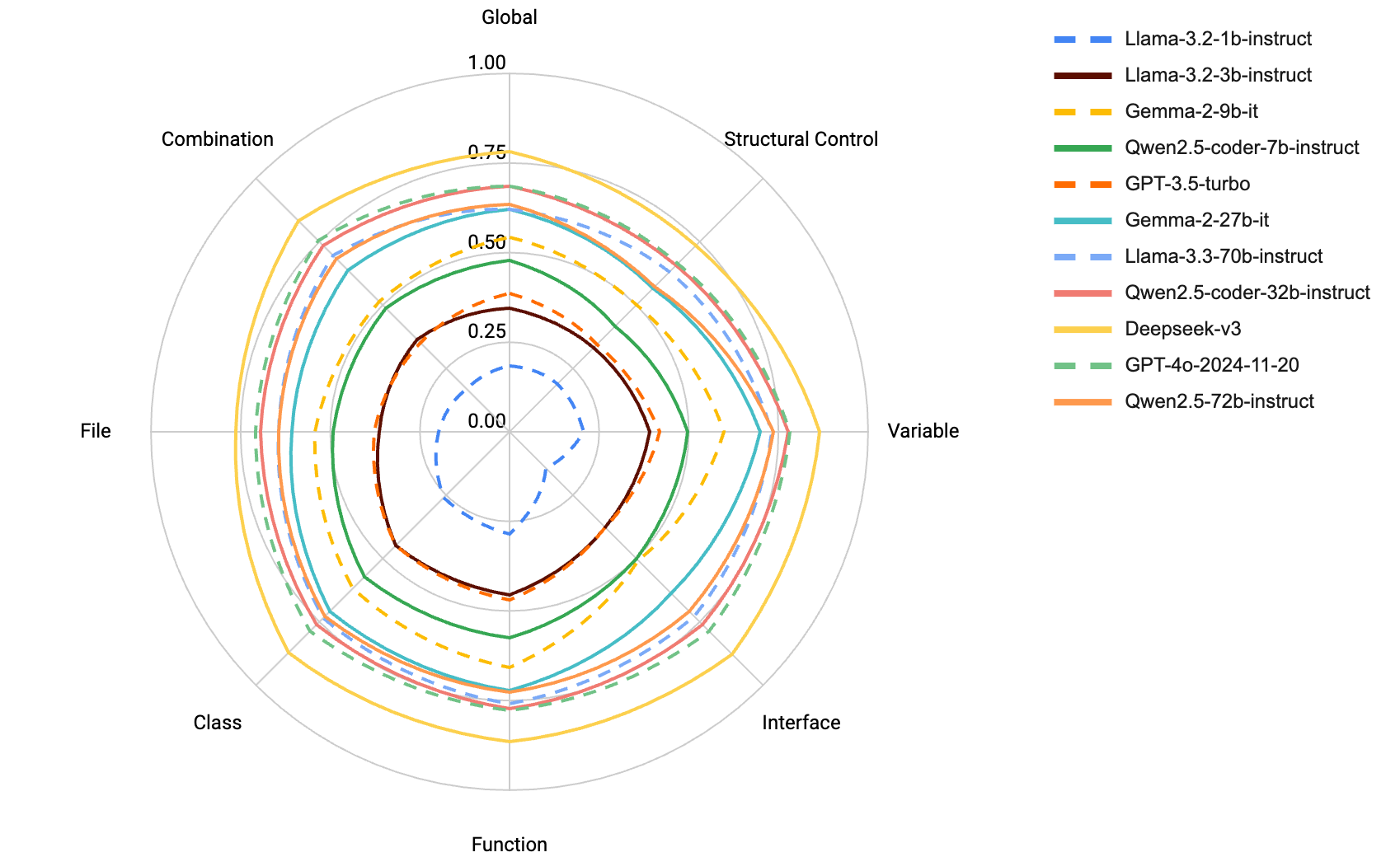}
    \caption{Performance of different LLMs on the CodeIF evaluation across instruction categories, measured by SSR.}
    \label{fig:model_result_instruct_type}
\end{figure}

\subsection{Performance Analysis Across Instruction Types}
Figure~\ref{fig:model_result_instruct_type} compares LLM performance across instruction categories, revealing notable variations. \textbf{DeepSeek-V3} leads overall, excelling in combination tasks (\textbf{0.831}) and global structure control, though weaker in variable handling, reflecting its optimization focus. \textbf{Meta's Llama series} shows a clear correlation between model size and performance, with larger variants (\textit{Llama-3.3-70B-Instruct}) outperforming smaller ones (\textit{Llama-3.2-1B-Instruct}). However, size alone is not decisive; comparisons with similarly sized models like \textbf{Google's Gemma} highlight the role of architecture and training methods in shaping performance.

\subsection{Cross-Language Performance Analysis of LLMs}

\begin{figure*}
    \centering
    \includegraphics[width=.75\textwidth]{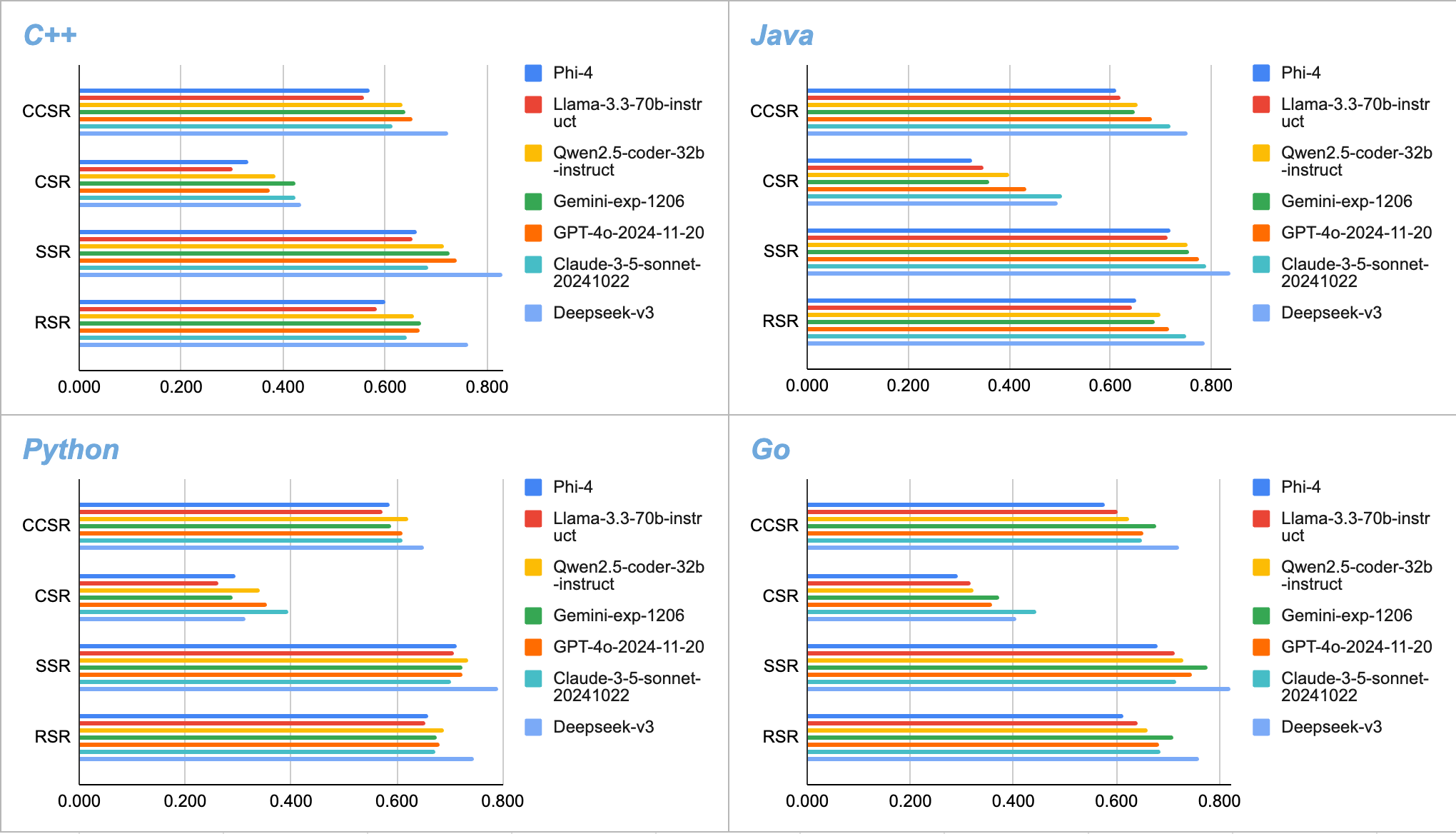}
    \caption{SSR scores of LLMs across different programming languages in the CodeIF evaluation.}
    \label{fig:model_result_lang}
\end{figure*}

Figure~\ref{fig:model_result_lang} compares the performance of leading LLMs across C++, Java, Python, and Go, highlighting trends at both model and language levels. At the \textbf{model level}, \textbf{DeepSeek-V3} leads with the highest CCSR in C++ (0.725), Java (0.753), and Go (0.722), and an RSR of 0.787 in Java. \textbf{Claude-3-5-Sonnet} excels in Java with the highest CSR (0.504) and RSR (0.749), but shows lower SSR in Python (0.703). \textbf{GPT-4O} demonstrates balanced performance, ranking second in Python's CSR (0.355) and RSR (0.682), with minimal variance (CV = 0.18). At the \textbf{language level}, C++ is the most challenging due to complex template metaprogramming. Java shows high inter-model variance, with Claude-3-5-Sonnet performing best. Go achieves the highest SSR but fluctuates in RSR. These results highlight cross-language generalization differences and suggest optimization areas like dependency handling and paradigm consistency.
\subsection{Analysis of Instruction Adherence Deviations}

Analysis of model-generated responses shows frequent deviations from instructions, especially in \textbf{naming conventions} and \textbf{formatting constraints}. Models often ignore global formatting rules, such as line limits, and inconsistently follow naming conventions. For example, when instructed to use \textbf{PascalCase}, models sometimes output lowercase or underscore-separated formats (e.g., incorrectly transforming \texttt{current\_power} into \texttt{CurrentPower}). A notable issue is the disregard for \textbf{prohibitive instructions}. For instance, models often use \texttt{if} statements despite being instructed to avoid them in favor of ternary operators or data structures like dictionaries, revealing gaps in constraint enforcement.

\subsection{Improving Instruction Compliance}  
Appendix Table~\ref{tab:codeIF_evaluation_all} highlights strategies to improve adherence. \textbf{Supervised Fine-Tuning (SFT)} proves effective, especially in the Llama series, while larger models like \texttt{Qwen2.5-72B-Instruct} outperform smaller ones in instruction-following accuracy. Key improvements include prioritizing \textit{hard constraints} (e.g., syntax rules) over \textit{soft guidelines} (e.g., coding styles). Patterned code generation can replace conditional statements with lookup tables or state machines. A naming convention engine can automate variable name formatting (e.g., converting \texttt{snake\_case} to \texttt{PascalCase}). \textit{Abstract Syntax Tree (AST)} analysis can detect and transform prohibited structures, such as replacing \texttt{for} loops with \texttt{while} loops. Conflict resolution mechanisms can address contradictory instructions, offering alternative solutions when certain language features are unavailable (e.g., using Python's alternatives to \texttt{switch-case}).

%% file: related_work.tex
\section{Related Work}

Code generation and instruction-following are pivotal capacities under examination in AI research~\cite{feng2020codebert,unicoder,wizardcoder,wang2023codet5+,kim2018deepcode,li2023starcoder,lu2021codexglue,li2022competition,wei2023magicoder,nijkamp2022codegen,zhuo2024bigcodebench,jain2024livecodebench,nijkamp2023codegen2,zhang2023repocoder,allal2023santacoder,lozhkov2024starcoder,roziere2023codellama,lozhkov2024starcoder2,wang-etal-2021-codet5,yan2023codetransocean}. Several benchmarks have been devised to appraise these capabilities in large-scale models. For code generation, benchmarks like McEval~\cite{mceval}, FullStackBench~\cite{liu2024fullstackbenchevaluatingllms}, Repocoder~\cite{zhang2023repocoder}, Repobench~\cite{liu2023repobench}, and LiveCodeBench~\cite{jain2024livecodebench} have been notable. Similarly, instruction-following capacities are gauged through benchmarks such as InfoBench~\cite{qin2024infobenchevaluatinginstructionfollowing}, CFBench~\cite{zhang2024cfbenchcomprehensiveconstraintsfollowingbenchmark}, Instruct-following~\cite{zhou2023instructionfollowingevaluationlargelanguage}, and FollowBench~\cite{jiang2024followbenchmultilevelfinegrainedconstraints}, each tailored to assess different aspects of following instructions given to models.

%% file: conclusion.tex
\section{Conclusion}

This study introduces \bench{}, a benchmark for evaluating the instruction-following capabilities of LLMs in code generation. Covering \textbf{Java, Python, Go, and C++}, CodeIF constructs a diverse test set with constraints ranging from global to specific variables. It introduces novel evaluation metrics—\textbf{Completely Satisfaction Rate (CSR), Soft Satisfaction Rate (SSR), Rigorous Satisfaction Rate (RSR), and Consistent Continuity Satisfaction Rate (CCSR)}—to assess multi-constraint handling across multiple dimensions.

%% file: limitations.tex
\section{Limitations}

\noindent \textbf{Limited Language Support.} CodeIF includes key languages like Java, Python, Go, and C++, but excludes popular ones like JavaScript, Ruby, and Swift. Expanding language coverage would improve its applicability in diverse contexts.

\noindent \textbf{Static Evaluation Focus.} CodeIF focuses mainly on static code properties, such as structure and naming conventions, while overlooking dynamic factors like runtime behavior, performance, and debugging. Including dynamic evaluation would better reflect real-world development challenges.

\noindent \textbf{Uniform Metric Weighting.} The metrics (CSR, SSR, RSR, CCSR) treat all constraints equally, which may not align with practical priorities. For example, syntactic correctness is often more critical than naming conventions. Introducing weighted scoring could enhance the interpretability of model performance.

%% file: appendix.tex
\section{Prompt Template} \label{appendix:prmpttemplate}

\begin{tcolorbox}[title=Prompt for Instruction Generation, label={fig:test_generation_prompt}]
You are an instruction compliance evaluator, required to assess the instruction compliance ability of large models. Therefore, you need to generate a series of data for the code generation instruction detection of large models.\\
\textbf{[Input Format]}\\
I will input a series of data, and you need to generate a dictionary based on these data, which includes two fields ``question'' and ``instruction\_list''\\
\textbf{Original question}:\\
\{Original question\}\\
\textbf{Original instruction list:}\\
\{instruction\_list\}\\
\textbf{Input Explanation}\\
The original question is the original question. It contains the original code generation problem.\\
The original instruction list is the original instruction list. It contains randomly generated code compliance instructions. Some instructions will contain directive keywords that need to be replaced and are wrapped in \{\{\}\}.

\textbf{Return Format}\\
Return a json data, do not have extra output.
The returned dictionary contains two fields: ``question'' and ``instruction\_list''\\
The format is as follows:
\begin{verbatim}
{
    "question": "Optimized question",
    "instruction_list": [
        {
            "instruction_id": "id1",
            "instruction": "Instruction 1"
        }
    ]
}
\end{verbatim}
\textbf{Explanation}\\
``question'': It is the optimized question, which does not contain any directive instructions, only contains the explanation of the original question. It does not contain any restrictions on the code. Move the instructions in the question to the instruction list\\
``instruction\_list'': It is the optimized instruction list. You should optimize according to the meaning of the question. More in line with the meaning of the question. Instead of directly outputting the original instruction list, note that you should replace all directive keywords that need to be replaced and are wrapped in {{}}, and the final output should not contain directive keywords that need to be replaced.\\
\textbf{Generation Requirements}\\
question: Please generate the optimized question based on the following data, which does not contain any directive instructions, only contains the core content of the original question.\\
instruction\_list: Originated from the input original instruction list. If there are instructions that completely conflict with the meaning of the question or instructions that conflict with each other \\
You should delete as little as possible, you should modify more.
Please replace according to the content in the original instruction\_list, you should delete as little as possible. Unless it is contradictory instructions, or instructions that cannot be achieved at all, if you only need to generate additional code to meet the requirements, you can replace it.
\end{tcolorbox}

\begin{tcolorbox}[title=Prompt for Code Generation, label={fig:test_generation_prompt}]
As a programming assistant, your task is to generate code snippets based on the user question and instructions given below:

\textbf{Please consider the following points while generating the code snippet}:

- Make sure you follow the user instructions word to word. If the instruction says to use a specific language or a specific method, use exactly that.

- Your output should be a valid code snippet in the programming language indicated in the question or the instructions.

- Pay close attention to the syntax and formatting rules of the programming language that you are using. The code should be well-formatted and easy to read. 

- Make sure that the code snippet you are generating is efficient and is not overly complicated.

\textbf{Output Format}:\\
The output should be a valid, well-formatted, and efficient code snippet that adheres to the above question and instructions.

\textbf{Task information}

User Question:\\
\{question\}

Instructions:\\
\{instructions\_str\}

Please generate the code snippet based on the above information:
\end{tcolorbox}

\begin{tcolorbox}[title=Prompt for  Answer Judgment]
\label{prompt:judge_prompt}
As a programming assistant, your task is to evaluate whether the generated code strictly follows the instructions given in light of the user's problem and directives. You need to return a list of the same length as the instructions, containing only 'Yes' and 'No', indicating whether the model adhered to each specific instruction.

\textbf{Consider the following when making judgments:}

- You must strictly follow the user's instructions. If the instruction requires the use of a specific language or method, you must explicitly check if the code utilizes it.

- Your output should be a list of the same length as the instructions, containing only 'Yes' and 'No'.

- Pay close attention to the programming language syntax and formatting rules you are evaluating. The code should be neatly organized and easy to read.

- The list you generate should be valid and not overly complex.

\textbf{Task Information}

\textbf{User question:}\\
\{question\}

\textbf{Instructions:}\\
\{instructions\_str\}

\textbf{Model-generated response:}\\
\{generated\_code\}

Based on the information provided, determine whether the model has followed the instructions, and return a list of the same length as the instructions, containing only `Yes' and `No'.
Please note!!! Your output should only contain the list, with no other content. The items in the list should only be `Yes' and `No', with no other words included.
\end{tcolorbox}

\section{More Resluts} \label{app:moreresults}
\begin{table}[h]
\centering
\small
\begin{tabular}{ccp{9cm}p{5cm}}
\toprule
ID & Type & Instruction Format & Format Keys \\
\midrule
1 & global & Your entire response should be written in \{programming\_language\}, the use of other programming languages is not allowed. & ["programming\_language"] \\
2 & global & Your code lines should not exceed \{characters\_num\} characters. & ["characters\_num"] \\
3 & global & Your code should use global variables. & [] \\
4 & global & Your code should not use global variables. & [] \\
5 & global & Your function should have at most \{parameter\_count\} parameters. & ["parameter\_count"] \\
6 & global & Your code should not have more than \{function\_count\} functions. & ["function\_count"] \\
7 & global & Your code should not have more than \{class\_count\} classes. & ["class\_count"] \\
8 & global & Your code should not use the \{keyword\} keyword. & ["keyword"] \\
9 & global & Your function should not exceed \{line\_num\} lines. & ["line\_num"] \\
10 & global & Your answer in total should not exceed \{line\_num\} lines. & ["line\_num"] \\
11 & global & Your code should use the \{keyword\} keyword. & ["keyword"] \\
12 & structural control & Your code should use data structure \{data\_structure\}. & ["data\_structure"] \\
13 & structural control & Your code should not use data structure \{data\_structure\}. & ["data\_structure"] \\
14 & structural control & Your code should use for-loop. & [] \\
15 & structural control & Your code should not use for-loop. & [] \\
16 & structural control & Your code should use while-loop. & [] \\
17 & structural control & Your code should not use while-loop. & [] \\
18 & structural control & Your code should use if statement for decision making. & [] \\
19 & structural control & Your code should not use if statement for decision making. & [] \\
20 & structural control & Your code should use switch statement for decision making. & [] \\
21 & structural control & Your code should not use switch statement for decision making. & [] \\
22 & variable & Your code should define a variable named \{variable\_name\}. & ["variable\_name"] \\
23 & variable & Your code should define an enumeration named \{enumeration\_name\} & ["enumeration\_name"] \\
24 & variable & The variable names in your code should follow the \{naming\_convention\} naming convention & ["naming\_convention"] \\
25 & variable & Variable \{variable\_name\}, type should be \{variable\_type\}. & ["variable\_name", "variable\_type"] \\
26 & variable & Variable \{variable\_name\}, should be a global variable. & ["variable\_name"] \\
27 & variable & Variable \{variable\_name\}, should not be a global variable. & ["variable\_name"] \\
28 & variable & Variable \{variable\_name\}, the initial value should be \{variable\_value\}. & ["variable\_name", "variable\_value"] \\
29 & variable & Variable \{variable\_name\}, should be a constant. & ["variable\_name"] \\
30 & variable & Variable \{variable\_name\} should not be a constant. & ["variable\_name"] \\
31 & function & Your code should include a function named \{function\_name\}. & ["function\_name"] \\
32 & function & The function names in your code should follow the \{naming\_convention\}. naming convention & ["naming\_convention"] \\
33 & function & Your code should not use any functions from the \{disallowed\_function\_list\}. & ["disallowed\_function\_list"] \\
34 & interface & Your code should define an interface named \{interface\_name\}. & ["interface\_name"] \\
35 & interface & The interface names in your code should follow the \{naming\_convention\} naming convention. & ["naming\_convention"] \\
36 & class & Your code should define a class named \{class\_name\}. & ["class\_name"] \\
37 & class & The class names in your code should follow the \{naming\_convention\} naming convention. & ["naming\_convention"] \\
38 & file & Your code should be organized in a package named \{package\_name\}. & ["package\_name"] \\
39 & file & Your code should import the following libraries \{library\_list\}. & ["library\_list"] \\
40 & file & Your code should use the function \{function\_name\} from the library \{library\_name\}. & ["function\_name", "library\_name"] \\
41 & file & Your code should not use the following libraries \{disallowed\_library\_list\}. & ["disallowed\_library\_list"] \\
42 & combination & You should initialize an object named \{object\_name\} as an instance of the \{class\_name\} class using \{parameters\_name\_list\} for initialization. & ["object\_name", "class\_name", "parameters\_name\_list"] \\
43 & combination & You should define an interface named \{interface\_name\} that includes these methods \{method\_name\_list\}. & ["interface\_name", "method\_name\_list"] \\
44 & combination & Your code should define a class named \{class\_name\} that implements the \{interface\_name\} interface. & ["class\_name", "interface\_name"] \\
45 & combination & In your code, the class \{class\_name\} should have these properties \{properties\_name\_list\}. & ["class\_name", "properties\_name\_list"] \\
46 & combination & In your code, the class \{class\_name\} should have these methods \{method\_name\_list\}. & ["class\_name", "method\_name\_list"] \\
47 & combination & The function \{function\_name\} should take \{parameter\_name\_list\} as parameters. & ["function\_name", "parameter\_name\_list"] \\
48 & combination & The function \{function\_name\} should return a \{return\_type\} as its result. & ["function\_name", "return\_type"] \\
49 & combination & Your code should be organized in a package named \{package\_name\}, which should contain these classes \{class\_name\_list\}. & ["package\_name", "class\_name\_list"] \\
50 & combination & Your code should be organized in a package named \{package\_name\}, which should contain these functions \{function\_name\_list\}. & ["package\_name", "function\_name\_list"] \\
\bottomrule
\end{tabular}
\caption{Constraint Instruction Table}
\label{tab:Constraint_Instruction_Tabl}
\end{table}

\begin{table*}[t]
\centering
\resizebox{1.0\textwidth}{!}{
\begin{tabular}{ccccccccccccc}
\toprule
\multirow{2}{*}{\textbf{Models}} & \multicolumn{3}{c}{\textbf{CSR}} & \multicolumn{3}{c}{\textbf{SSR}} & \multicolumn{3}{c}{\textbf{RSR}} & \multicolumn{3}{c}{\textbf{CCSR}} \\
\cmidrule(lr){2-13} 
\textbf{} & \textbf{Full} & \textbf{Easy} & \textbf{Hard} & \textbf{Full} & \textbf{Easy} & \textbf{Hard} & \textbf{Full} & \textbf{Easy} & \textbf{Hard} & \textbf{Full} & \textbf{Easy} & \textbf{Hard} \\
\midrule
\textbf{Llama-3.2-1b-instruct} & 0.034 & 0.046 & 0.022 & 0.218 & 0.277 & 0.159 & 0.182 & 0.231 & 0.132 & 0.152 & 0.197 & 0.107\\
\textbf{Qwen2.5-1.5b-instruct} & 0.034 & 0.053 & 0.015 & 0.265 & 0.334 & 0.197 & 0.222 & 0.282 & 0.162 & 0.181 & 0.234 & 0.128\\
\textbf{Qwen2.5-coder-1.5b-instruct} & 0.058 & 0.086 & 0.03 & 0.358 & 0.436 & 0.281 & 0.301 & 0.371 & 0.233 & 0.251 & 0.314 & 0.189\\
\textbf{Qwen2.5-3b-instruct} & 0.078 & 0.109 & 0.046 & 0.415 & 0.489 & 0.34 & 0.357 & 0.432 & 0.282 & 0.299 & 0.364 & 0.233\\
\textbf{Llama-3.2-3b-instruct} & 0.101 & 0.137 & 0.065 & 0.396 & 0.473 & 0.318 & 0.344 & 0.419 & 0.268 & 0.305 & 0.375 & 0.235\\
\textbf{GPT-3.5-turbo} & 0.102 & 0.14 & 0.065 & 0.41 & 0.467 & 0.353 & 0.362 & 0.42 & 0.303 & 0.314 & 0.369 & 0.259\\
\textbf{Qwen2.5-coder-3b-instruct} & 0.097 & 0.142 & 0.051 & 0.445 & 0.529 & 0.359 & 0.383 & 0.464 & 0.301 & 0.33 & 0.401 & 0.258\\
\textbf{Llama-3.1-8b} & 0.129 & 0.178 & 0.08 & 0.452 & 0.551 & 0.353 & 0.402 & 0.497 & 0.306 & 0.352 & 0.44 & 0.263\\
\textbf{Llama-3.1-8b-instruct} & 0.145 & 0.187 & 0.102 & 0.467 & 0.544 & 0.388 & 0.418 & 0.493 & 0.34 & 0.37 & 0.444 & 0.295\\
\textbf{Qwen2.5-coder-7b-instruct} & 0.142 & 0.198 & 0.087 & 0.514 & 0.59 & 0.438 & 0.453 & 0.533 & 0.373 & 0.39 & 0.463 & 0.318\\
\textbf{Ministral-3b} & 0.127 & 0.162 & 0.092 & 0.526 & 0.591 & 0.46 & 0.458 & 0.527 & 0.39 & 0.4 & 0.458 & 0.342\\
\textbf{Phi-3.5-mini-128k-instruct} & 0.154 & 0.217 & 0.09 & 0.514 & 0.635 & 0.391 & 0.456 & 0.574 & 0.337 & 0.405 & 0.51 & 0.299\\
\textbf{Qwen2.5-7b-instruct} & 0.153 & 0.201 & 0.104 & 0.535 & 0.599 & 0.471 & 0.475 & 0.546 & 0.405 & 0.416 & 0.479 & 0.353\\
\textbf{Ministral-8b} & 0.161 & 0.205 & 0.116 & 0.552 & 0.614 & 0.489 & 0.486 & 0.552 & 0.419 & 0.431 & 0.49 & 0.371\\
\textbf{Gemma-2-9b-it} & 0.171 & 0.21 & 0.131 & 0.573 & 0.642 & 0.504 & 0.513 & 0.587 & 0.44 & 0.445 & 0.508 & 0.383\\
\textbf{Llama-3.1-70b} & 0.196 & 0.232 & 0.16 & 0.61 & 0.664 & 0.555 & 0.545 & 0.607 & 0.482 & 0.482 & 0.533 & 0.43\\
\textbf{Qwen2.5-coder-14b-instruct} & 0.218 & 0.276 & 0.16 & 0.596 & 0.667 & 0.525 & 0.539 & 0.614 & 0.463 & 0.483 & 0.55 & 0.416\\
\textbf{Qwen2.5-14b-instruct} & 0.238 & 0.279 & 0.198 & 0.61 & 0.676 & 0.543 & 0.557 & 0.628 & 0.486 & 0.498 & 0.565 & 0.431\\
\textbf{Gemini-2.0-flash-exp} & 0.254 & 0.29 & 0.218 & 0.615 & 0.648 & 0.583 & 0.556 & 0.593 & 0.518 & 0.514 & 0.547 & 0.481\\
\textbf{Gemma-2-27b-it} & 0.245 & 0.3 & 0.19 & 0.658 & 0.709 & 0.607 & 0.596 & 0.652 & 0.54 & 0.533 & 0.588 & 0.478\\
\textbf{Llama-3.1-70b-instruct} & 0.265 & 0.3 & 0.229 & 0.675 & 0.723 & 0.627 & 0.612 & 0.667 & 0.556 & 0.559 & 0.601 & 0.516\\
\textbf{Qwen2.5-32b-instruct} & 0.294 & 0.337 & 0.251 & 0.68 & 0.722 & 0.638 & 0.621 & 0.674 & 0.568 & 0.56 & 0.604 & 0.515\\
\textbf{Qwen2.5-72b-instruct} & 0.281 & 0.319 & 0.244 & 0.685 & 0.734 & 0.634 & 0.621 & 0.677 & 0.564 & 0.569 & 0.619 & 0.518\\
\textbf{Codestral-2501} & 0.28 & 0.339 & 0.219 & 0.683 & 0.748 & 0.617 & 0.621 & 0.691 & 0.551 & 0.571 & 0.633 & 0.507\\
\textbf{Phi-4} & 0.312 & 0.361 & 0.262 & 0.698 & 0.735 & 0.66 & 0.635 & 0.681 & 0.589 & 0.589 & 0.631 & 0.546\\
\textbf{Llama-3.3-70b-instruct} & 0.307 & 0.359 & 0.255 & 0.698 & 0.749 & 0.647 & 0.632 & 0.691 & 0.574 & 0.589 & 0.643 & 0.536\\
\textbf{GPT-4o-mini} & 0.292 & 0.348 & 0.237 & 0.731 & 0.78 & 0.682 & 0.665 & 0.724 & 0.606 & 0.609 & 0.66 & 0.559\\
\textbf{GPT-4o} & 0.338 & 0.392 & 0.283 & 0.721 & 0.77 & 0.671 & 0.665 & 0.721 & 0.609 & 0.616 & 0.668 & 0.563\\
\textbf{Qwen2.5-coder-32b-instruct} & 0.365 & 0.422 & 0.307 & 0.736 & 0.767 & 0.704 & 0.679 & 0.723 & 0.635 & 0.634 & 0.669 & 0.599\\
\textbf{Gemini-exp-1206} & 0.357 & 0.41 & 0.303 & 0.744 & 0.781 & 0.707 & 0.685 & 0.734 & 0.636 & 0.636 & 0.675 & 0.597\\
\textbf{Gemini-1.5-pro} & 0.351 & 0.383 & 0.318 & 0.763 & 0.794 & 0.732 & 0.704 & 0.744 & 0.663 & 0.647 & 0.679 & 0.615\\
\textbf{GPT-4o-2024-11-20} & 0.383 & 0.441 & 0.325 & 0.748 & 0.792 & 0.702 & 0.689 & 0.745 & 0.633 & 0.65 & 0.698 & 0.602\\
\textbf{Claude-3-5-sonnet-20241022} & \textbf{0.444} & \textbf{0.525} & 0.362 & 0.727 & 0.784 & 0.669 & 0.692 & 0.757 & 0.626 & 0.652 & 0.715 & 0.587\\
\textbf{Deepseek-coder} & 0.41 & 0.45 & \textbf{0.37} & 0.805 & 0.836 & 0.773 & 0.749 & 0.791 & 0.707 & 0.699 & 0.732 & 0.666\\
\textbf{Deepseek-v3} & 0.414 & 0.468 & 0.359 & \textbf{0.821} & \textbf{0.847} & \textbf{0.794} & \textbf{0.764} & \textbf{0.806} & \textbf{0.723} & \textbf{0.712} & \textbf{0.743} & \textbf{0.68}\\

\bottomrule
\end{tabular}}
\caption{ CodeIF evaluation results of different difficulties. We use bold font to mark the best results in all models.}
\label{tab:codeIF_evaluation_all}
\end{table*}

\begin{table}[h]
\centering
\small
\resizebox{1.0\textwidth}{!}{
\begin{tabular}{c|cccccccc|}
\toprule
\textbf{Models} &\textbf{Global} &\textbf{Structural Control}&\textbf{Variable} &\textbf{Interface} &\textbf{Function} &\textbf{Class} &\textbf{File} &\textbf{Combination}\\
\midrule

\textbf{Llama-3.2-1b-instruct} & 0.186 & 0.190 & 0.206 & 0.144 & 0.284 & 0.260 & 0.198 & 0.172 \\
\textbf{Qwen2.5-1.5b-instruct} & 0.244 & 0.236 & 0.221 & 0.213 & 0.355 & 0.315 & 0.230 & 0.213 \\
\textbf{Qwen2.5-coder-1.5b-instruct} & 0.328 & 0.304 & 0.326 & 0.293 & 0.436 & 0.426 & 0.351 & 0.304 \\
\textbf{Qwen2.5-3b-instruct} & 0.383 & 0.346 & 0.412 & 0.383 & 0.468 & 0.481 & 0.383 & 0.366 \\
\textbf{Llama-3.2-3b-instruct} & 0.344 & 0.332 & 0.393 & 0.376 & 0.454 & 0.447 & 0.363 & 0.367 \\
\textbf{GPT-3.5-turbo} & 0.388 & 0.344 & 0.417 & 0.375 & 0.467 & 0.449 & 0.378 & 0.352 \\
\textbf{Qwen2.5-coder-3b-instruct} & 0.397 & 0.367 & 0.438 & 0.419 & 0.511 & 0.507 & 0.415 & 0.403 \\
\textbf{Llama-3.1-8b} & 0.410 & 0.355 & 0.451 & 0.424 & 0.500 & 0.503 & 0.413 & 0.413 \\
\textbf{Llama-3.1-8b-instruct} & 0.422 & 0.373 & 0.482 & 0.455 & 0.524 & 0.499 & 0.407 & 0.437 \\
\textbf{Qwen2.5-coder-7b-instruct} & 0.479 & 0.419 & 0.497 & 0.502 & 0.576 & 0.571 & 0.492 & 0.487 \\
\textbf{Ministral-3b} & 0.472 & 0.403 & 0.527 & 0.512 & 0.618 & 0.609 & 0.524 & 0.535 \\
\textbf{Phi-3.5-mini-128k-instruct} & 0.461 & 0.410 & 0.512 & 0.531 & 0.562 & 0.574 & 0.485 & 0.491 \\
\textbf{Qwen2.5-7b-instruct} & 0.484 & 0.425 & 0.532 & 0.548 & 0.616 & 0.591 & 0.520 & 0.520 \\
\textbf{Ministral-8b} & 0.497 & 0.433 & 0.541 & 0.570 & 0.622 & 0.631 & 0.527 & 0.557 \\
\textbf{Gemma-2-9b-it} & 0.541 & 0.498 & 0.599 & 0.510 & 0.659 & 0.618 & 0.543 & 0.511 \\
\textbf{Llama-3.1-70b} & 0.558 & 0.500 & 0.652 & 0.653 & 0.685 & 0.671 & 0.545 & 0.597 \\
\textbf{Qwen2.5-coder-14b-instruct} & 0.541 & 0.467 & 0.623 & 0.669 & 0.645 & 0.652 & 0.547 & 0.594 \\
\textbf{Qwen2.5-14b-instruct} & 0.569 & 0.526 & 0.652 & 0.592 & 0.649 & 0.644 & 0.533 & 0.559 \\
\textbf{Gemini-2.0-flash-exp} & 0.555 & 0.526 & 0.653 & 0.666 & 0.685 & 0.669 & 0.564 & 0.615 \\
\textbf{Gemma-2-27b-it} & 0.621 & 0.569 & 0.699 & 0.640 & 0.722 & 0.710 & 0.607 & 0.637 \\
\textbf{Llama-3.1-70b-instruct} & 0.606 & 0.546 & 0.722 & 0.718 & 0.744 & 0.738 & 0.603 & 0.680 \\
\textbf{Qwen2.5-32b-instruct} & 0.637 & 0.581 & 0.713 & 0.712 & 0.732 & 0.742 & 0.601 & 0.653 \\
\textbf{Qwen2.5-72b-instruct} & 0.633 & 0.570 & 0.734 & 0.711 & 0.727 & 0.726 & 0.645 & 0.686 \\
\textbf{Codestral-2501} & 0.617 & 0.552 & 0.723 & 0.718 & 0.733 & 0.746 & 0.651 & 0.694 \\
\textbf{Phi-4} & 0.633 & 0.586 & 0.734 & 0.739 & 0.721 & 0.752 & 0.677 & 0.710 \\
\textbf{Llama-3.3-70b-instruct} & 0.621 & 0.634 & 0.733 & 0.730 & 0.759 & 0.738 & 0.645 & 0.695 \\
\textbf{GPT-4o-mini} & 0.671 & 0.663 & 0.787 & 0.774 & 0.784 & 0.783 & 0.657 & 0.710 \\
\textbf{GPT-4o} & 0.665 & 0.651 & 0.742 & 0.759 & 0.743 & 0.759 & 0.666 & 0.716 \\
\textbf{Qwen2.5-coder-32b-instruct} & 0.683 & 0.654 & 0.776 & 0.763 & 0.772 & 0.758 & 0.695 & 0.736 \\
\textbf{Gemini-exp-1206} & 0.690 & 0.677 & 0.780 & 0.789 & 0.798 & 0.809 & 0.675 & 0.727 \\
\textbf{Gemini-1.5-pro} & 0.718 & 0.696 & 0.814 & 0.800 & 0.812 & 0.815 & 0.672 & 0.749 \\
\textbf{GPT-4o-2024-11-20} & 0.685 & 0.666 & 0.784 & 0.786 & 0.779 & 0.785 & 0.706 & 0.755 \\
\textbf{Claude-3-5-sonnet-20241022} & 0.677 & 0.678 & 0.750 & 0.736 & 0.742 & 0.730 & 0.640 & 0.692 \\
\textbf{Deepseek-coder} & 0.759 & 0.714 & 0.850 & 0.856 & 0.846 & 0.847 & 0.754 & 0.813 \\
\textbf{Deepseek-v3} & 0.780 & 0.732 & 0.866 & 0.876 & 0.866 & 0.873 & 0.762 & 0.831 \\

\bottomrule
\end{tabular}}
\caption{The performance of various models on CodeIF for different types of instructions}
\end{table}

\begin{table*}[t]
\small
\centering
\resizebox{1.0\textwidth}{!}{
\begin{tabular}{c|cccc|cccc|cccc|cccc}
\toprule

\multirow{2}{*}{\textbf{Models}} & \multicolumn{4}{c}{\textbf{CPP}} & \multicolumn{4}{c}{\textbf{Java}} & \multicolumn{4}{c}{\textbf{Python}} & \multicolumn{4}{c}{\textbf{Go}} \\
\cmidrule(lr){2-17} 
\textbf{} & \textbf{CCS}&\textbf{CS} & \textbf{SS} & \textbf{RS} & \textbf{CCS} & \textbf{CS} & \textbf{SS} & \textbf{RS}& \textbf{CCS} & \textbf{CS} & \textbf{SS} & \textbf{RS}& \textbf{CCS} & \textbf{CS} & \textbf{SS} & \textbf{RS} \\
\midrule

\textbf{Llama-3.2-1b-instruct} & 0.123 & 0.023 & 0.185 & 0.150 & 0.190 & 0.037 & 0.265 & 0.221 & 0.179 & 0.047 & 0.262 & 0.223 & 0.086 & 0.022 & 0.117 & 0.096 \\
\textbf{Qwen2.5-1.5b-instruct} & 0.171 & 0.023 & 0.250 & 0.206 & 0.191 & 0.026 & 0.277 & 0.228 & 0.197 & 0.047 & 0.298 & 0.257 & 0.151 & 0.040 & 0.216 & 0.179 \\
\textbf{Qwen2.5-coder-1.5b-instruct} & 0.253 & 0.068 & 0.348 & 0.297 & 0.259 & 0.055 & 0.375 & 0.308 & 0.263 & 0.060 & 0.380 & 0.328 & 0.218 & 0.049 & 0.309 & 0.255 \\
\textbf{Qwen2.5-3b-instruct} & 0.251 & 0.046 & 0.367 & 0.302 & 0.310 & 0.078 & 0.419 & 0.367 & 0.306 & 0.092 & 0.435 & 0.384 & 0.327 & 0.093 & 0.433 & 0.365 \\
\textbf{Llama-3.2-3b-instruct} & 0.284 & 0.073 & 0.377 & 0.313 & 0.345 & 0.121 & 0.435 & 0.380 & 0.321 & 0.112 & 0.429 & 0.383 & 0.244 & 0.084 & 0.304 & 0.265 \\
\textbf{GPT-3.5-turbo} & 0.301 & 0.085 & 0.388 & 0.332 & 0.367 & 0.134 & 0.461 & 0.409 & 0.265 & 0.092 & 0.371 & 0.334 & 0.318 & 0.088 & 0.412 & 0.363 \\
\textbf{Qwen2.5-coder-3b-instruct} & 0.339 & 0.103 & 0.444 & 0.380 & 0.338 & 0.101 & 0.453 & 0.391 & 0.320 & 0.091 & 0.446 & 0.390 & 0.323 & 0.093 & 0.431 & 0.363 \\
\textbf{Llama-3.1-8b} & 0.319 & 0.115 & 0.409 & 0.354 & 0.366 & 0.130 & 0.477 & 0.420 & 0.376 & 0.152 & 0.485 & 0.446 & 0.330 & 0.110 & 0.413 & 0.363 \\
\textbf{Llama-3.1-8b-instruct} & 0.328 & 0.112 & 0.432 & 0.375 & 0.408 & 0.173 & 0.503 & 0.447 & 0.393 & 0.147 & 0.496 & 0.455 & 0.325 & 0.133 & 0.406 & 0.365 \\
\textbf{Qwen2.5-coder-7b-instruct} & 0.389 & 0.147 & 0.505 & 0.434 & 0.375 & 0.118 & 0.503 & 0.444 & 0.400 & 0.155 & 0.531 & 0.475 & 0.401 & 0.154 & 0.516 & 0.456 \\
\textbf{Ministral-3b} & 0.356 & 0.107 & 0.473 & 0.401 & 0.410 & 0.150 & 0.542 & 0.476 & 0.404 & 0.112 & 0.538 & 0.481 & 0.430 & 0.138 & 0.544 & 0.464 \\
\textbf{Phi-3.5-mini-128k-instruct} & 0.354 & 0.108 & 0.461 & 0.388 & 0.426 & 0.179 & 0.532 & 0.478 & 0.440 & 0.180 & 0.559 & 0.510 & 0.380 & 0.131 & 0.482 & 0.422 \\
\textbf{Qwen2.5-7b-instruct} & 0.401 & 0.162 & 0.514 & 0.448 & 0.439 & 0.152 & 0.559 & 0.495 & 0.397 & 0.147 & 0.523 & 0.471 & 0.429 & 0.154 & 0.541 & 0.485 \\
\textbf{Ministral-8b} & 0.400 & 0.143 & 0.518 & 0.439 & 0.434 & 0.158 & 0.560 & 0.495 & 0.410 & 0.142 & 0.538 & 0.481 & 0.494 & 0.214 & 0.599 & 0.532 \\
\textbf{Gemma-2-9b-it} & 0.446 & 0.200 & 0.560 & 0.499 & 0.446 & 0.164 & 0.576 & 0.518 & 0.380 & 0.131 & 0.510 & 0.456 & 0.542 & 0.204 & 0.678 & 0.609 \\
\textbf{Llama-3.1-70b} & 0.487 & 0.201 & 0.598 & 0.518 & 0.507 & 0.232 & 0.632 & 0.572 & 0.425 & 0.136 & 0.579 & 0.521 & 0.522 & 0.226 & 0.635 & 0.571 \\
\textbf{Qwen2.5-coder-14b-instruct} & 0.464 & 0.224 & 0.572 & 0.514 & 0.478 & 0.206 & 0.592 & 0.535 & 0.522 & 0.216 & 0.653 & 0.594 & 0.454 & 0.235 & 0.544 & 0.490 \\
\textbf{Qwen2.5-14b-instruct} & 0.481 & 0.230 & 0.590 & 0.533 & 0.528 & 0.265 & 0.639 & 0.581 & 0.472 & 0.188 & 0.599 & 0.550 & 0.511 & 0.281 & 0.603 & 0.557 \\
\textbf{Gemini-2.0-flash-exp} & 0.491 & 0.259 & 0.587 & 0.519 & 0.575 & 0.309 & 0.664 & 0.604 & 0.468 & 0.207 & 0.584 & 0.533 & 0.514 & 0.233 & 0.619 & 0.558 \\
\textbf{Gemma-2-27b-it} & 0.529 & 0.271 & 0.645 & 0.579 & 0.551 & 0.261 & 0.676 & 0.616 & 0.465 & 0.179 & 0.604 & 0.543 & 0.611 & 0.289 & 0.727 & 0.665 \\
\textbf{Llama-3.1-70b-instruct} & 0.535 & 0.267 & 0.653 & 0.578 & 0.581 & 0.276 & 0.685 & 0.620 & 0.555 & 0.251 & 0.696 & 0.639 & 0.557 & 0.264 & 0.655 & 0.596 \\
\textbf{Qwen2.5-32b-instruct} & 0.551 & 0.314 & 0.655 & 0.602 & 0.589 & 0.330 & 0.706 & 0.638 & 0.522 & 0.231 & 0.665 & 0.609 & 0.580 & 0.311 & 0.690 & 0.634 \\
\textbf{Qwen2.5-72b-instruct} & 0.543 & 0.297 & 0.638 & 0.574 & 0.580 & 0.288 & 0.701 & 0.633 & 0.574 & 0.284 & 0.702 & 0.651 & 0.573 & 0.249 & 0.687 & 0.610 \\
\textbf{Codestral-2501} & 0.562 & 0.307 & 0.658 & 0.595 & 0.583 & 0.301 & 0.694 & 0.632 & 0.566 & 0.249 & 0.693 & 0.637 & 0.569 & 0.261 & 0.681 & 0.611 \\
\textbf{Phi-4} & 0.570 & 0.331 & 0.663 & 0.601 & 0.612 & 0.328 & 0.719 & 0.650 & 0.587 & 0.295 & 0.714 & 0.660 & 0.577 & 0.292 & 0.679 & 0.613 \\
\textbf{Llama-3.3-70b-instruct} & 0.558 & 0.300 & 0.652 & 0.582 & 0.621 & 0.348 & 0.713 & 0.644 & 0.572 & 0.264 & 0.709 & 0.653 & 0.602 & 0.317 & 0.712 & 0.640 \\
\textbf{GPT-4o-mini} & 0.582 & 0.292 & 0.684 & 0.615 & 0.620 & 0.299 & 0.738 & 0.667 & 0.586 & 0.261 & 0.731 & 0.674 & 0.661 & 0.330 & 0.775 & 0.707 \\
\textbf{GPT-4o} & 0.600 & 0.337 & 0.698 & 0.639 & 0.652 & 0.368 & 0.748 & 0.693 & 0.600 & 0.312 & 0.723 & 0.676 & 0.603 & 0.332 & 0.701 & 0.636 \\
\textbf{Qwen2.5-coder-32b-instruct} & 0.633 & 0.384 & 0.717 & 0.658 & 0.654 & 0.401 & 0.753 & 0.699 & 0.621 & 0.342 & 0.736 & 0.688 & 0.624 & 0.322 & 0.731 & 0.661 \\
\textbf{Gemini-exp-1206} & 0.640 & 0.424 & 0.726 & 0.672 & 0.650 & 0.360 & 0.755 & 0.689 & 0.590 & 0.290 & 0.724 & 0.674 & 0.677 & 0.373 & 0.777 & 0.710 \\
\textbf{Gemini-1.5-pro} & 0.635 & 0.370 & 0.741 & 0.676 & 0.674 & 0.379 & 0.783 & 0.720 & 0.610 & 0.278 & 0.758 & 0.706 & 0.674 & 0.395 & 0.764 & 0.707 \\
\textbf{GPT-4o-2024-11-20} & 0.653 & 0.374 & 0.741 & 0.669 & 0.683 & 0.434 & 0.776 & 0.716 & 0.612 & 0.355 & 0.724 & 0.682 & 0.653 & 0.358 & 0.747 & 0.683 \\
\textbf{Claude-3-5-sonnet-20241022} & 0.615 & 0.425 & 0.684 & 0.643 & 0.720 & 0.504 & 0.789 & 0.749 & 0.611 & 0.396 & 0.703 & 0.674 & 0.650 & 0.444 & 0.716 & 0.686 \\
\textbf{Deepseek-coder} & 0.709 & 0.441 & 0.802 & 0.735 & 0.731 & 0.463 & 0.819 & 0.764 & 0.657 & 0.336 & 0.791 & 0.747 & 0.702 & 0.403 & 0.805 & 0.744 \\
\textbf{Deepseek-v3} & 0.725 & 0.435 & 0.831 & 0.762 & 0.753 & 0.497 & 0.839 & 0.787 & 0.651 & 0.315 & 0.793 & 0.744 & 0.722 & 0.404 & 0.822 & 0.76 \\

\bottomrule
\end{tabular}}
\caption{ the evaluation results of different languages on CODEIF. The metrics include Consistent Continuity Satisfaction Rate (CCSR), Complete Satisfaction Rate (CSR), Soft Satisfaction Rate (SSR), and Rigorous Satisfaction Rate (RSR).}
\label{tab:multitool-normal}
\end{table*}
\clearpage

\section{Baselines} \label{app:baselines}
We evaluate over 30 language models spanning both open-source architectures and commercial APIs. The Meta Llama 3 Series~\cite{touvron2023llama} contains \textit{Llama-3.2-1B/3B/8B/70B-Instruct} variants and \textit{Llama-3.3-70B-Instruct}. Qwen2.5 Series~\cite{yang2024qwen2} encompasses \textit{Qwen2.5-1.5B/3B/7B/14B/32B/72B-Instruct} with dedicated code generation models \textit{Qwen2.5-Coder-1.5B/3B/7B/14B/32B-Instruct}~\cite{hui2024qwen25codertechnicalreport}. Mistral Series~\cite{Jiang2023Mistral7} includes \textit{Mistral-3B}, \textit{Mistral-8B}, and the code-specialized \textit{Codestral-2501}.

The evaluation covers Microsoft's \textit{Phi-3.5-Mini-128K-Instruct (3.8B)} and \textit{Phi-4}~\cite{abdin2024phi4technicalreport}, along with Google's \textit{Gemma-2-9B/27B-It}~\cite{gemmateam2024gemmaopenmodelsbased}. DeepSeek Series incorporates \textit{DeepSeek-Coder}~\cite{guo2024deepseek} and \textit{DeepSeek-V3}~\cite{deepseekai2024deepseekv3technicalreport}. Commercial APIs include OpenAI's \textit{GPT-3.5-Turbo}, \textit{GPT-4O-Mini}, \textit{GPT-4O-2024-05-13}, and \textit{GPT-4O-2024-11-20}~\cite{achiam2023gpt4}; Google's \textit{Gemini-2.0-Flash-Exp}, \textit{Gemini-Exp-1206}, and \textit{Gemini-1.5-Pro}~\cite{geminiteam2024geminifamilyhighlycapable}; plus Anthropic's \textit{Claude-3.5-Sonnet-20241022}.

\section{More Data Analysis}  \label{app:dataanalysis}

\begin{figure}
    \centering
    \includegraphics[width=.7\linewidth]{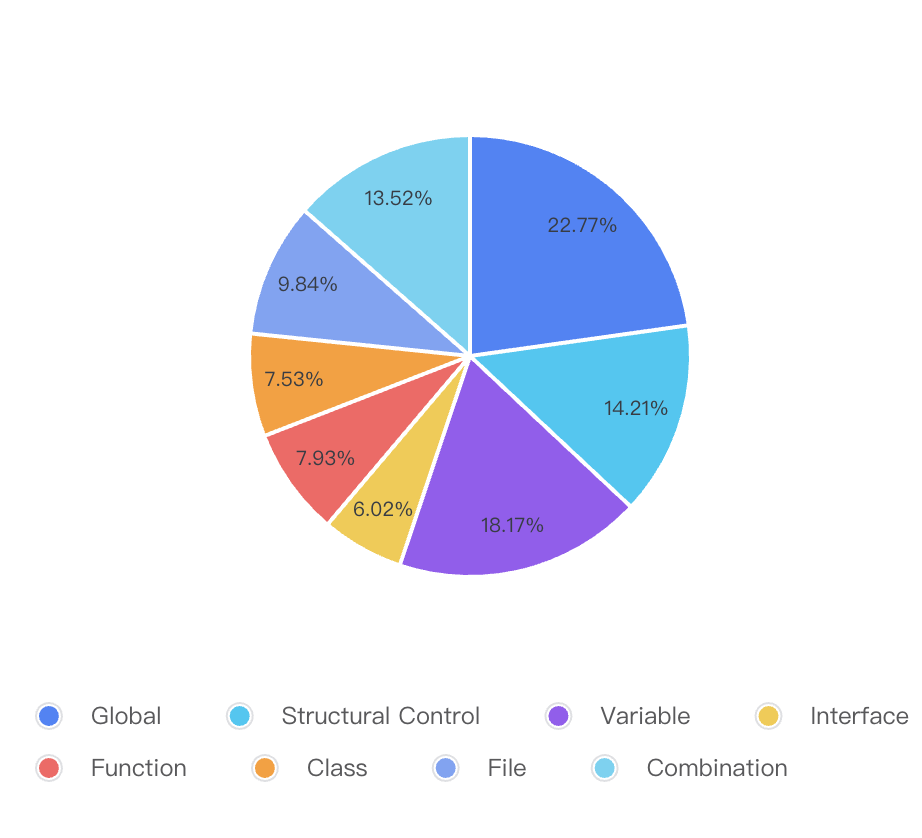}
    \caption{Distribution of atomic instruction list lengths across difficulty levels.}
    \label{fig:data_set_type}
\end{figure}

\begin{figure}
    \centering
    \includegraphics[width=.5\textwidth]{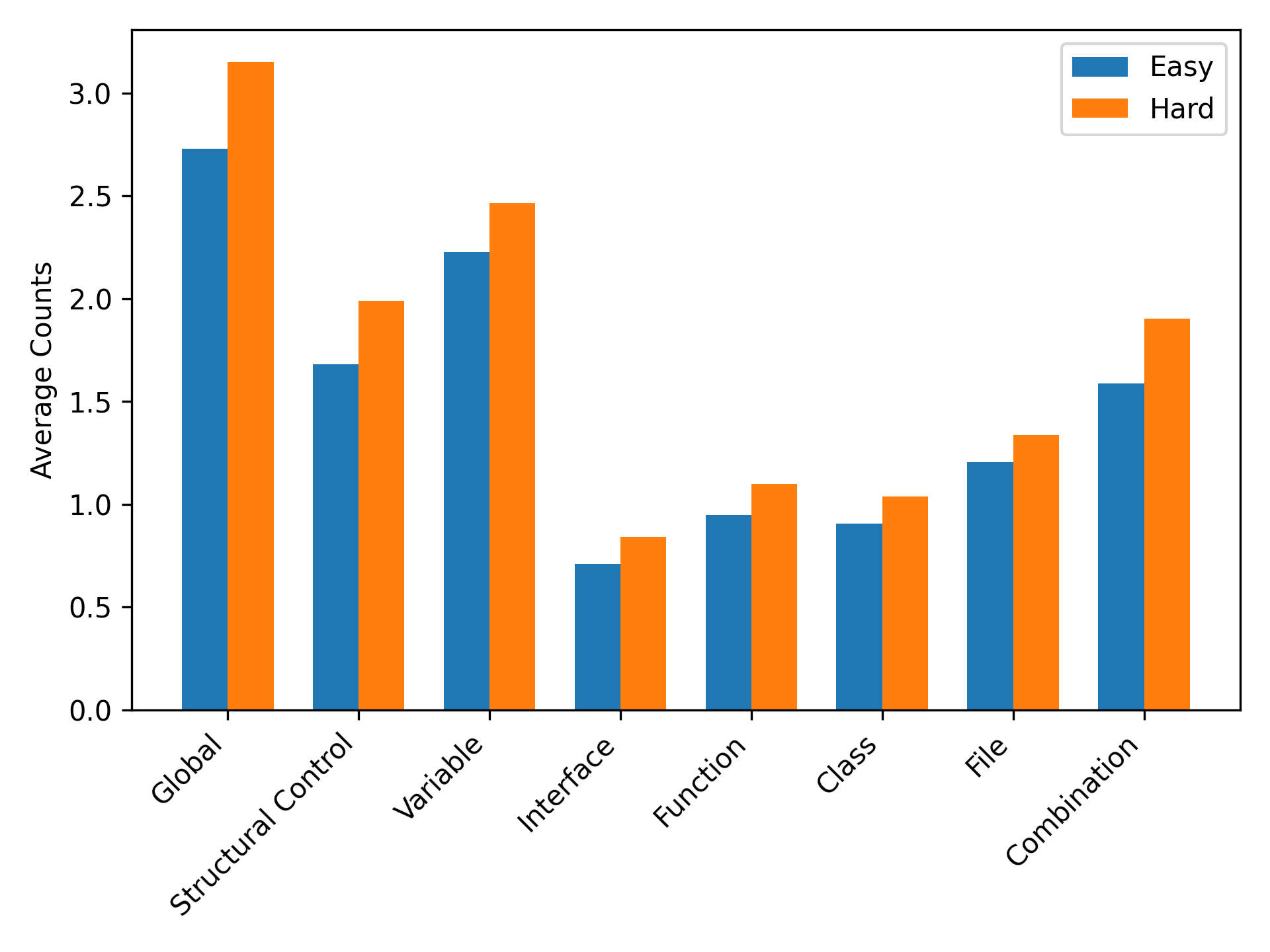}
    \caption{ The distribution of constraint instruction list lengths in datasets of different difficulties.}
    \label{fig:data_set_instruct_type}
\end{figure}

Figure~\ref{fig:data_set_type} shows the proportion of each instruction category. \textbf{Global} constraints dominate (\textbf{22.77\%}), followed by \textbf{Variable} constraints (\textbf{18.17\%}). This distribution reflects \textbf{CodeIF}’s balanced focus on high-level structural coherence and fine-grained variable precision, ensuring comprehensive evaluation of code generation capabilities. Figure~\ref{fig:data_set_instruct_type} compares instruction distribution across difficulty levels.

